\documentclass[%
 reprint,
 superscriptaddress,
 amsmath,amssymb,
 aps,
 pre,
 onecolumn,
]{revtex4-2}

\usepackage{graphicx}
\usepackage{dcolumn}
\usepackage{bm}
\usepackage{upgreek}
\usepackage{bm}
\usepackage{xcolor}
\usepackage{booktabs}
\usepackage{cellspace}%
\usepackage{siunitx}
\usepackage{makecell}
\usepackage{hyperref}
\setcellgapes{3pt}

\begin{document}

\preprint{APS/123-QED}

\title{Diffusion from particle-coated drops: the subtle role of particle size }

\author{Alexandros T. Oratis}
\affiliation{Delft University of Technology, Department of Chemical Engineering, van der Maasweg 9, 2629 HZ Delft, The Netherlands}

\author{Matteo Camagna}
\affiliation{Delft University of Technology, Department of Chemical Engineering, van der Maasweg 9, 2629 HZ Delft, The Netherlands}

\author{Timo J.J.M. van Overveld}
\affiliation{Delft University of Technology, Department of Chemical Engineering, van der Maasweg 9, 2629 HZ Delft, The Netherlands}

\author{Valeria Garbin}
\affiliation{Delft University of Technology, Department of Chemical Engineering, van der Maasweg 9, 2629 HZ Delft, The Netherlands}

\date{\today}

\begin{abstract}
Many natural and industrial systems involve particle-laden interfaces. 
Because interfacial particles prevent the coalescence and coarsening of drops, they hold promise for various applications requiring stable emulsions. 
Despite their remarkable ability to stabilize emulsions, it remains challenging to characterize how particles influence the interfacial transport of dissolved solutes. 
Here, we quantify the diffusion from a single particle-coated drop by confining it to a two-dimensional configuration.
Using fluorescence microscopy, we extract the intensity profiles of the fluorescent dye as it diffuses from the drop, yielding spatio-temporal measurements of the concentration field.
Over a range of particle sizes, the particles impose minimal resistance to diffusion. 
We rationalize this counterintuitive result with a mathematical model that couples interfacial mass transfer to a particle-coated interface. 
We show that the particle monolayer controls the temporal dynamics of the flux across the interface, hindering transport only at extreme coverage fractions beyond the close-packing limit. 
This framework reveals why particles often fail to hinder diffusion, offering new pathways to harness mass transfer in particle-stabilized emulsions.
\end{abstract}

\maketitle



\section{Introduction}
Complex fluid interfaces involving colloidal particles appear in a variety of natural and industrial systems \cite{hunter2008role,sagis2011dynamic,tang2015stimuli}.
By adhering to an interface, particles reduce the surface free energy \cite{binks2002particles}, altering the functional response of the liquid interface in terms of its rheology and rigidity \cite{
van2017interfacial,thijssen2018interfacial}.
In the context of drops and bubbles, interfacial particles prevent coalescence and coarsening, leading to remarkable stability of emulsions and foams \cite{pawar2011arrested,chevalier2013emulsions,cates2018theories}.
The enhanced stability provided by particles has been exploited for a broad class of applications that require long-lasting emulsions, ranging from cosmetic products \cite{bais2005rheological,cates2018theories} and processed foods \cite{berton2015pickering,yan2020protein}, to chemical conversion \citep{crossley2010solid,rodriguez2020catalysis,chang2021recent,dedovets2022multiphase}.
An important physical mechanism that dictates the performance of these systems involves the transport of solutes across the liquid interface. 
In food and cosmetic products it is advantageous to prevent diffusion, as dissolution and ripening of drops and bubbles can diminish the desired functional and sensory properties.
In contrast, diffusion is critical for the exchange of reactants and products in chemical conversion \cite{rodriguez2020catalysis} and drug release \cite{dinsmore2002colloidosomes,thompson2015colloidosomes,sun2019coated}.
It is thus imperative to fundamentally understand how the presence of particles at the liquid interface affects the diffusion dynamics.

When particles adhere to interfaces, they reduce the overall surface area available for mass transport.
When the particles are densely packed at the liquid interface, the pores through which solutes may diffuse become fewer or smaller in size.
Because the diffusive flux is typically proportional to the surface area, this reasoning suggests that particles will impede the transport of solutes.
Indeed, various experimental studies demonstrate that particles can dramatically affect diffusion-governed processes.
For instance, particles will arrest the capillary-driven dissolution of bubbles \cite{abkarian2007dissolution} or form multilayers that slow down the evaporation of liquid marbles \cite{laborie2013coatings,gallo2021particle,saczek2024impact}.
Yet, these two model systems involve retracting interfaces, which significantly affect the diffusion rates by reducing the interfacial area.
Conversely, in systems involving the diffusive exchange of solutes from drops with constant sizes, the effect of the particle layer remains ambiguous.
While various studies report a strong hindrance on mass transport in the presence of a particle coating \cite{kim2007colloidal,sjoo2015barrier,li2025pickering}, others demonstrate a negligible effect \cite{dinsmore2002colloidosomes,lee2008double,yow2009release,thompson2010covalently,liu2024diffusion}. 
These conflicting results highlight the need for systematically quantifying the effects of particle-coated interfaces on diffusion. 

Our recent theoretical investigations on diffusion through planar interfaces coated with spherical particles demonstrated that the particles did not significantly influence mass transport \cite{liu2024diffusion,van2025scaling}.
The reduction in interfacial area available for diffusion induced by the particles does not limit the transport of dissolved solutes, unless the coverage fraction is increased to extreme values beyond the close-packing limit of spheres.
A simple criterion based on the surface coverage and particle size can predict when the coating hinders diffusion, which is consistent with the experimental results reported in the literature \cite{van2025scaling}.
Yet none of these experimental measurements were performed in model configurations to systematically examine the effects of the particle monolayer.
Furthermore, the theoretical predictions are limited to transport between liquids with identical transport and thermodynamic properties.
Therefore, a generalized and validated framework for diffusion from particle-coated drops is still lacking.

In this study, we aim to quantify and generalize the diffusion dynamics from an isolated particle-coated drop into a surrounding liquid.
We perform fluorescent microscopy experiments to characterize the diffusion dynamics across a wide range of particle sizes.
We characterize the effects of the particles by measuring the radial extent to which the fluorescent dye penetrates the outer phase, as well as the time it takes to fully deplete the drop.
In addition to the experiments, we perform finite-difference simulations on the diffusion across particle-laden interfaces.
Our theoretical approach couples the coating, transport, and thermodynamic properties, whose collective effects can be captured by two dimensionless parameters.
Our combined framework demonstrates that particle size has a subtle effect on the transport dynamics. 


\section{Methods}
\subsection{Water-in-oil emulsions}

Our experimental protocol closely follows the one previously adopted by our group \cite{liu2024diffusion}, based on fluorescent microscopy imaging.
We first generated coated drops by preparing particle-stabilized emulsions, consisting of water drops surrounded by hexadecane.
We dispersed sulfate latex polystyrene (PS) beads (Invitrogen, Thermo Fisher Scientific) in water at a concentration of 2\% w/v using ultrasonication for 5 minutes.
The suspension was then mixed by equal volumes with a 2 mM Rhodamine B (RhB) solution, which acts as the fluorescent tracer to track the concentration field.
A small amount of a salt ($50$ mM $\rm CaCl_2$) was added to screen the electrostatic repulsion between the negatively charged polystyrene particles and promote their adsorption to the liquid interface during emulsification.
The suspension was deposited in a vial with hexadecane at a volume ratio of 20:80 v/v.
The mixture was emulsified using a vortex mixer for two minutes, leading to a stable emulsion of particle-coated water drops, whose sizes ranged from 100 to 1000 $\SI{}{\micro\meter}$.
We note that the negative charge of the PS particles interacts with the positively charged group of the RhB fluorescent dye, causing the dye to adhere on the surface of the particles. 
Therefore, only a fraction of the initial RhB concentration remains within the coated water drops, whose initial concentrations $c_0$ can be estimated a posteriori by computing the total amount of dye diffused from the drop.

\subsection{Confined two-dimensional drop geometry}
\begin{figure*}[t!]
    \centering
    \includegraphics[width=\linewidth]{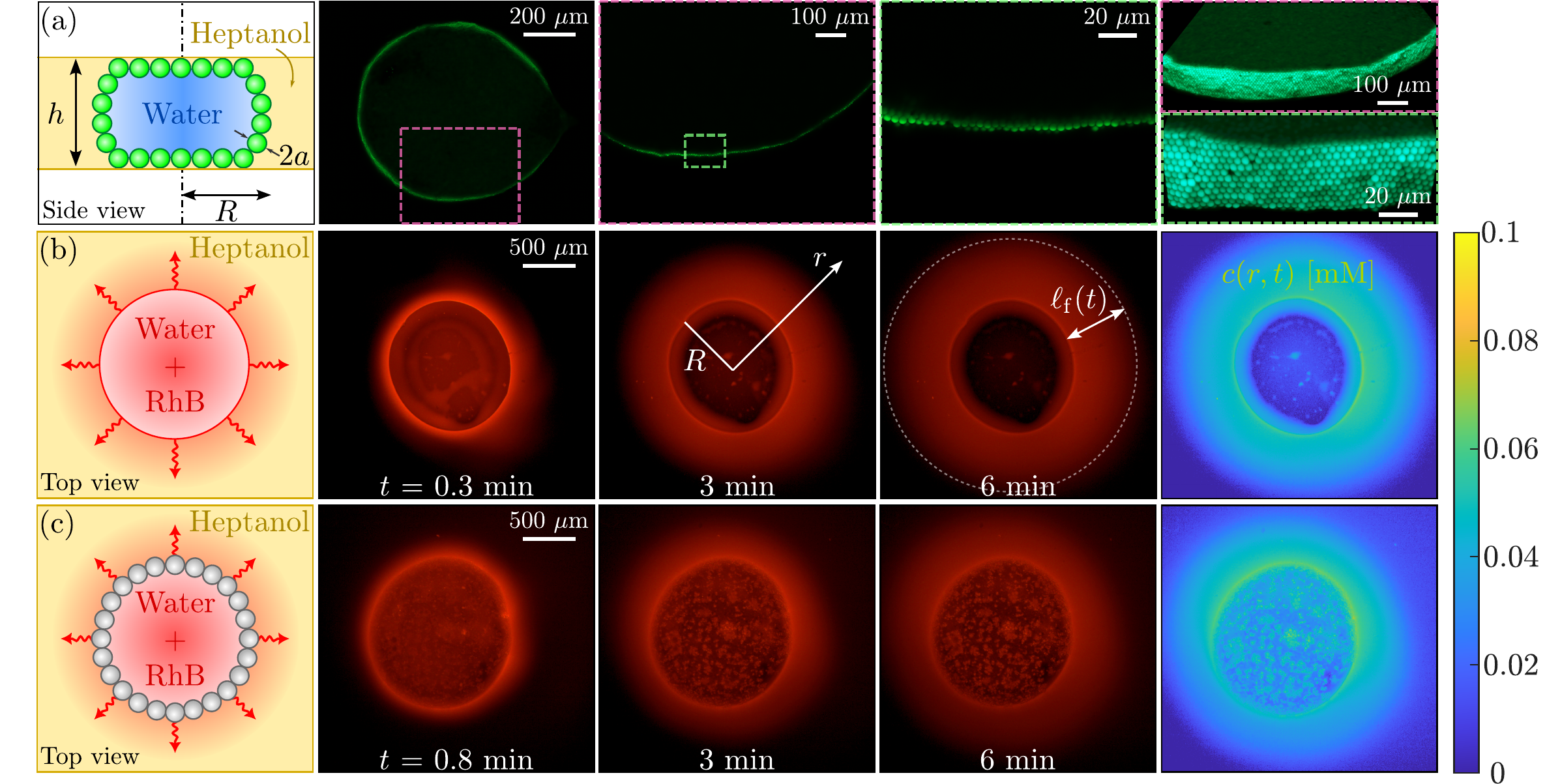}
    \caption{Experimental snapshots of the diffusion from a drop with radius $R = 0.6$ mm.
    (a) Images of a drop coated with fluorescent polystyrene beads (radius $a = 2$ $\SI{}{\micro\meter}$) obtained using confocal microscopy at different magnifications.
    The last panel illustrates the three-dimensional reconstruction of the particle coating.
    (b) The diffusion dynamics can be visualized using fluorescence microscopy by dissolving Rhodamine B (RhB) inside the drop.
    For a bare drop surrounded by heptanol, the dye diffuses radially outwards within minutes.
     The characteristic radial length scale quantifying the penetration depth is denoted as $\ell_{\rm f}(t)$, which grows with time.
    (c) Coating the drop with polystyrene particles (radius $a = 5$ $\SI{}{\micro\meter}$) still leads to the diffusion of RhB towards the outer phase. 
    The image intensities can be converted to concentration fields $c(r,t)$ measured in [mM].}
    \label{Fig1}
\end{figure*}

To study the diffusion in a controlled manner, 
we isolate a particle-coated drop and confine it between two glass slides using an image spacer (Grace bio-Labs, Secure-Seal, radius 13 mm and thickness $h = 175$ $\SI{}{\micro\meter}$).
After the coated drop is deposited in the spacer, the spacer is filled with 1-Heptanol.
A cover slip is then gently placed to cover the image spacer, squeezing the drop to a cylindrical pancake-like shape (Fig.~\ref{Fig1}(a)).
The resulting two-dimensional droplet radius varies from $R = 0.40$ mm to 0.85 mm. 
The confined geometry allows for precise control of transport phenomena, and has been successfully used to study droplet evaporation 
\cite{clement2004evaporation,boulogne2013buckling,loussert2016drying,milark2026confined}.
To systematically test the effect of the particle size, experiments were conducted for different particle sizes, with the particle radius $a = 0.05,\,0.25,\,0.50,\,0.90,\,2.50,\,5.00$ $\mu$m spanning two orders of magnitude.

Squeezing the coated drop from a spherical geometry to a pancake-like shape can alter the morphology of the particle coating.
To better understand how the change in geometry affects the coating morphology, we used confocal imaging to visualize the interfacial particles.
We prepared Pickering emulsions stabilized by fluorescent polystyrene particles (Invitrogen, Thermo Fisher Scientific $a = 2$ $\mu$m), from which a single drop was squeezed to a pancake shape.
Figure~\ref{Fig1}(a) presents snapshots of the particle layer at different magnifications, showing the monolayer of particles at the interface between the drop and the surrounding liquid.
The three-dimensional reconstructions of the coating illustrate that the particles can maintain a hexagonal configuration.
From these images, the coverage fraction is estimated to lie between $\phi_0 = 0.65$ and 0.75.

\subsection{Fluorescence microscopy}

To visualize the diffusion process taking place in the two-dimensional cell we used a fluorescence microscopy setup.
The RhB dye was excited with a wavelength of 540 nm, resulting in a fluorescent signal with emission wavelength of 625 nm, whose spatio-temporal evolution was recorded using a digital camera (DCC1645C, Thorlabs, frame rate of 0.1 fps) mounted on a microscope (Olympus BXFM, 2$\times$ magnification).
Typical snapshots of the diffusion process from a drop with radius $R = 0.6$ mm are shown in Figs.~\ref{Fig1}(b) and \ref{Fig1}(c) for a bare drop and a coated drop with $a = 5$ $\SI{}{\micro\meter}$, respectively.
We point out that because diffusion occurs immediately once heptanol contacts the drop, the time delay caused by placing the cover slip prevents us from capturing the dynamics at the very early stages.

We chose water-heptanol-RhB for the multiphase-multicomponent system, as RhB emits a strong fluorescent signal in both liquids, which is vital to obtain accurate concentration measurements.
In addition, RhB is soluble in both liquids and approximately 90 times more soluble in heptanol than in water.
Specifically, the partition coefficient of RhB in water and heptanol was measured by bringing water-RhB solution in contact with a heptanol bath (1:1 v/v).
After allowing the dye to equilibrate for two days, the partition coefficient was measured using a spectrophotometer (Hach Lange DR5000) and found to be $\kappa \equiv c_{\rm d}/c_{\rm b} = 1/88$, where $c_{\rm d}$ and $c_{\rm b}$ are the concentrations of the water drop and heptanol bath, respectively.
The diffusivity of RhB in water and heptanol was computed using the Stokes-Einstein relation \cite{bruus2007theoretical}, yielding $D_{\rm d} \approx 4.2\times 10^{-10}$ $\rm m^2/s$ and $D_{\rm b} \approx 0.6\times 10^{-10}$ $\rm m^2/s$ respectively.

The captured image sequences were post-processed to obtain concentration fields.
First, the raw images were converted to grayscale by extracting the intensity of the red channel $G(r,t)$, which varies with time $t$ and radial position $r$ from the center of the drop.
The intensity in arbitrary units [a.u.] was converted to concentration values in [mM] by performing separate calibration experiments using known concentrations.
The calibration was linear, such that the concentration field $c(r,t)$ can be computed as $c(r,t) = b G(r,t)$, with a calibration factor $b = 0.14\,{\rm mM/a.u.}$ and $ b = 0.18 \,{\rm mM/a.u.}$ for heptanol and water respectively. 
The linear calibration fit remained valid provided that the normalized intensities did not exceed values of 70\% that could arise from saturated images.
Therefore, all of our experiments were conducted under the exact same illumination settings and exposure times to allow for the consistent conversion to concentration scales.
Because the dye concentration  $c(r,t)$ adopts a radially symmetric profile, we compute the radial dependence by taking the average along the circumference at every radial position $r$ from the center of the drop.

\begin{figure*}[t!]
    \centering
    \includegraphics[width=\linewidth]{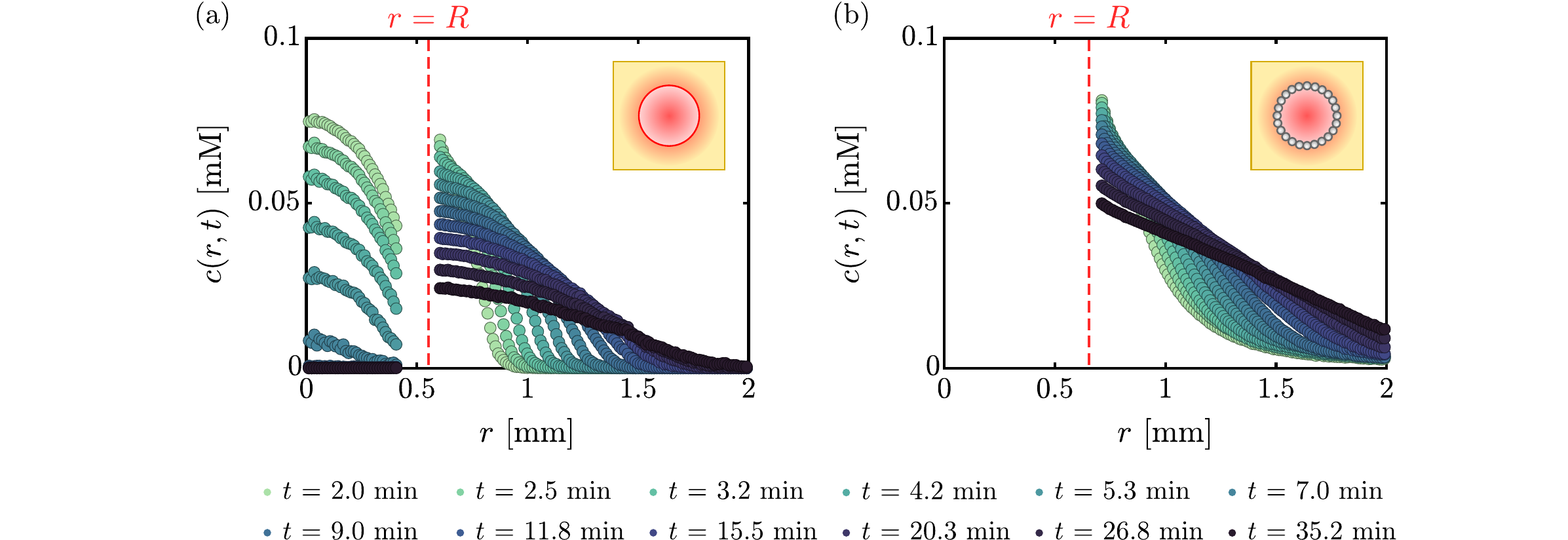}
    \caption{Spatio-temporal evolution of the concentration field $c(r,t)$.
    (a) The concentration inside the bare drop decreases radially outwards.
    Crossing the interface, the favorable solubility of RhB towards heptanol causes a jump in the concentration, which then decreases radially.
 As time progresses, the concentration within the drop decreases and falls to zero once the drop has been depleted.
 Meanwhile, the concentration spreads radially outward in the heptanol bath.
    (b)  Similar trends can be observed for the concentration field in heptanol generated by a coated drop with $a = 5$ $\SI{}{\micro\meter}$. }
    \label{Fig2}
\end{figure*}

\section{Measurements of concentration fields}

The spatiotemporal evolution of the concentration field  $c(r,t)$ is shown in Fig.~\ref{Fig2}, where it is plotted against the radial position $r$ for different time instances.
We start with an uncoated drop of radius $R = 0.6$ mm and initial RhB concentration $c_0 \approx 0.1$ mM, corresponding to the snapshots shown in Fig.~\ref{Fig1}(b).
During the early stages of diffusion the dye is still within the drop.
The concentration profile is fairly flat near the center and sharply decreases close to the interface (Fig.~\ref{Fig2}(a)).
The jump in concentration expected from the favorable solubility of RhB towards heptanol can be clearly seen in the concentration profile as we cross the interface at $r = R$.
As time progresses, the concentration profile spreads radially outward, while the interfacial concentration overall decreases.
Similarly, the concentration within the drop decreases over time, maintaining the same spatial features of a radially decreasing profile.
After approximately 7 minutes, the dye has diffused entirely into the outer phase.

We now switch to the concentration field generated by a coated drop with particle size $a = 5$ $\SI{}{\micro\meter}$ (Fig.~\ref{Fig2}(b)).
The concentration within the drop is no longer accessible, as the particle coating prevents an accurate measurement of the spatial variation of the intensity.
At each time instant, the concentration smoothly decreases in the radial direction, with the concentration gradient at the interface appearing to be sharper compared to the bare drop.
As the dye penetrates into the bath, the interfacial concentration decreases over time.
The overall trends of the spatio-temporal evolution of the concentration fields thus appear to be similar for the bare and coated drops.
However, the concentration fields alone are insufficient to evaluate how the coating affects the diffusion dynamics.

\begin{figure*}[b]
    \centering
    \includegraphics[width=\linewidth]{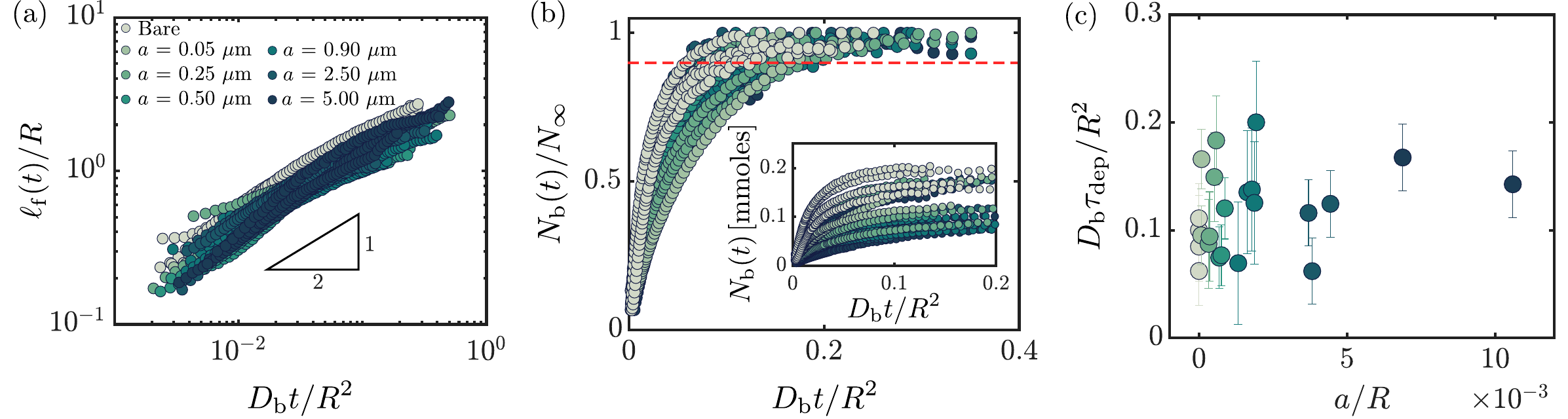}
    \caption{Effect of particle size on the diffusion dynamics. 
    (a) The scaled front length $\ell_{\rm f}/R$ plotted against dimensionless time $D_{\rm b}t/R^2$ for different particle radii $a$. 
    The curves fall on top of each other, following the scaling  $\ell_{\rm f} \sim t^{1/2}$.
    (b) Amount of diffused material $N_{\rm b}(t)$ normalized by its equilibrium value $N_\infty$ against the dimensionless time $D_{\rm b}t/R^2$. 
    The collapse of the data suggests a negligible effect of the particles on how much dye diffuses into the bath.
    The red dashed line corresponds to when $N_{\rm b}(t) = 0.9 N_\infty$, which is used to compute the depletion time.
   Inset: The dimensional amount of diffused material in [mmoles] plotted against the dimensionless time. 
   (c) The normalized depletion time $\tau_{\rm dep} D_{\rm b}/R^2$ remains fairly constant with the normalized particle size $a/R$. The error bars are estimated by computing the depletion time as $N_{\rm b}(t) =  0.95 N_\infty$.}
    \label{Fig3}
\end{figure*}

One way to assess the influence of the particle coating is to quantify how far the dye penetrates into the bath.
To this end, we compute the penetration length $\ell_{\rm f}(t)$, labeled in Fig.~\ref{Fig1}(b), which is computed as the radial distance at which the normalized intensity $G$ falls below 10\%.
The penetration length scaled by the drop radius $R$ is plotted against the dimensionless time $D_{\rm b}t/R^2$ (Fig.~\ref{Fig3}(a)).
We overlay the data for the different particle sizes $a$, which fall on top of that of a bare drop.
Each data set follows a power law $\ell_{\rm f} \sim (D_{\rm b} t)^{1/2}$, often encountered in diffusive boundary layers, even in two-dimensional circular geometries \cite{penas2017diffusion}.
Therefore, over a wide range of particle sizes, the coating does not affect the evolution of $\ell_{\rm f}(t)$. 
Yet, even if the penetration depth remains unaltered by the presence of the particles, the amount of dye that has escaped the drop may be significantly reduced.
We can thus re-examine the particle effects by investigating the time at which the total mass of the dye has diffused into the outer phase.

The total mass of diffused RhB in the outer phase can be computed by integrating the concentration field as $N_{\rm b}(t) = \int_{R}^\infty 2\pi r h c(r,t) {\rm d}r$.
Plotting $N_{\rm b}(t)$ against the dimensionless time $D_{\rm b} t/R^2$, we observe a rapid increase at short times, before it plateaus to a constant equilibrium value $N_{\infty}$ (Fig.~\ref{Fig3}(b) inset).
The equilibrium amount $N_{\infty}$ varies between each experimental run, as it depends on the initial concentration $c_0$ inside the drop.
Because the initial drop concentration is challenging to quantify for coated drops, we can assess the effects of the particles by measuring instead the time at which $N_{\rm b}(t)$ plateaus. 
We thus compute the depletion time $\tau_{\rm dep}$ by measuring the time at which the total mass $N_{\rm b}(t)$ reaches 90\% of $N_\infty$ (dashed line in Fig.~\ref{Fig3}(b)).
Plotting the dimensionless depletion time $D_{\rm b} \tau_{\rm dep}/R^2$ against the particle size relative to the drop radius $a/R$, we find a variability in the data without any clear trend (Fig.~\ref{Fig3}(c)).
The scatter of the depletion time can arise from variability in the surface coverage which does not remain constant between each experimental run.
Nevertheless, since both the penetration length and depletion time do not exhibit any correlation with the particle coating, it is reasonable to conclude that their presence has a negligible effect on the diffusion dynamics.

\section{Diffusion across a particle-coated interface}

\begin{figure*}[t]
    \centering
    \includegraphics[width=\linewidth]{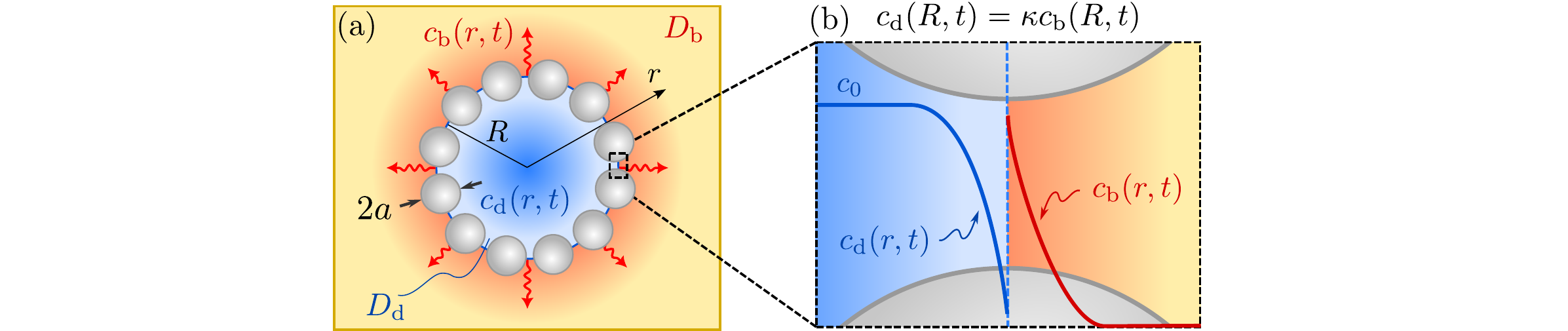}
    \caption{(a) Schematic illustrating the diffusion from a drop with radius $R$ and initial concentration $c_0$. 
    The drop has a concentration $c_{\rm d}(r,t)$ and diffusivity $D_{\rm d}$, while the surrounding bulk liquid has a concentration $c_{\rm b}(r,t)$ and diffusivity $D_{\rm b}$.
    (b) At the onset of diffusion, equilibrium partitioning at the interface relates the interfacial concentrations between the two phases, governed by the partition coefficient $\kappa$. }
    \label{Fig4}
\end{figure*}

\subsection{Model description}
To better understand why the particles have minimal effect on the transport dynamics, we develop a one-dimensional diffusion model for a pancake drop with a particle-coated interface.
We extend previous models on diffusion through a coated planar interface \cite{liu2024diffusion,van2025scaling}, which demonstrated that the particles only influence the dynamics at early times.
We generalize this framework to a two-dimensional drop using a cylindrical coordinate system, with $r = 0$ defined at the center of the drop.
The concentration field $c(r,t)$ is assumed to be axisymmetric and can be decomposed into a drop $c_{\rm d}(r,t)$ and bulk $c_{\rm b}(r,t)$ concentration (Fig.~\ref{Fig4}(a)).
The interfacial particles reduce the available area for diffusion, whose effect on the diffusion dynamics can be incorporated in the Fick-Jacobs equation \cite{jacobs1967diffusion,zwanzig1992diffusion}
\begin{equation}
\frac{\partial c_{\rm d}}{\partial t} = \frac{D_{\rm d}}{A(r)}\frac{1}{r}\frac{\partial}{\partial r}\left(r A(r) \frac{\partial c_{\rm d}}{\partial r} \right)\quad {\rm for} \quad 0 < r < R,
\label{eq:diffusion_w}
\end{equation}
and 
\begin{equation}
\frac{\partial c_{\rm b}}{\partial t} = \frac{D_{\rm b}}{A(r)}\frac{1}{r}\frac{\partial}{\partial r}\left(r A(r) \frac{\partial c_{\rm b}}{\partial r} \right)\quad {\rm for} \quad R < r < \infty.
\label{eq:diffusion_o}
\end{equation}
Here, $A(r) = A_0 [1 - \phi(r) ]$ is the interfacial area available for diffusion, which depends on the uncoated area $A_0$, and the particle surface coverage $\phi(r)$ that varies spatially across the layer.
Finally, we account for the steep variations of the available area near the particle edges by introducing spatially varying diffusivity corrections \cite{van2025scaling,reguera2001kinetic,kalinay2006corrections}.

Quantifying how the particles are arranged on the drop's interface is non-trivial, particularly when the spherically coated drop is squeezed into a disk-like shape.
Indeed, the curvature of the pancake drop limits the amount of spheres that can fit on the interface separating the two liquids.
While the particles can be arranged on a planar interface with a well-defined surface coverage $\phi_0$, for a cylindrical drop the coverage becomes dependent on the ratio $a/R$.
To minimize the complexity of our model, we relax this constraint and allow the particles to be arranged in lattices that are typically associated with planar interfaces.
This quasi-1D model for the coating oversimplifies the particle arrangement on the interface, but should provide a fairly accurate description for $a/R \ll 1$.
We thus proceed by assuming that the particles form a monolayer, whose surface coverage varies spatially as
\begin{equation}
\phi(r)  = 
	\begin{cases}
	\phi_0 \left[1 - \left(\displaystyle\frac{r - R}{a}\right)^2 \right] & {\rm for} \quad |r - R| \leq  a,\\[6pt]
	0 & {\rm for} \quad |r - R| >  a.
	\end{cases}
	\label{eq:phi}
\end{equation} 
To fully understand the effects of the surface coverage, we  explore a wider range of $\phi_0$ than accessed experimentally.
We note that because we consider a two-dimensional cylindrical geometry, both the concentrations $c_{\rm d}$ and $c_{\rm b}$, as well as the area $A(r)$, are expressed per arbitrary unit depth $h$.

The two diffusion equations are complemented by a symmetry boundary condition at the center of the drop $c_{\rm d}'(0,t) = 0$ and a vanishing far-field concentration in the bulk $c_{\rm b}(\infty,t) = 0$. 
Here, the prime denotes a differentiation with respect to $r$.
At the interface, equilibrium partitioning of the solute relates the drop and bulk concentrations $c_{\rm d}(R,t) = \kappa c_{\rm b}(R,t)$, as illustrated in Fig.~\ref{Fig4}(b).
In addition, the continuity of the flux at the interface sets $D_{\rm d} c_{\rm d}'(R,t) = D_{\rm b} c_{\rm b}'(R,t)$.
Finally, the initial condition for each concentration is $c_{\rm d}(r,0) = c_0$ and $c_{\rm b}(r,0) = 0$.
We numerically solve equations \eqref{eq:diffusion_w} and \eqref{eq:diffusion_o} using a centered finite difference scheme with logarithmically-spaced grid points and time steps --- see \cite{van2025scaling} for more details.
We first non-dimensionalize the radial distance $r$ by the drop radius $R$ and time $t$ by the bulk's diffusive time scale $R^2/D_{\rm b}$. 
The concentrations in the drop and the bulk are scaled by $c_0$ and $c_0/\kappa$ respectively.

The non-dimensionalization introduces four dimensionless parameters, reflecting the rich nature of the problem.
These parameters can be separated into two groups: first, the mass transfer and thermodynamic properties give rise to the diffusivity ratio $\mathcal{D} = D_{\rm b}/D_{\rm d}$ and partition coefficient $\kappa$ respectively. 
Second, the particle coating introduces the surface coverage $\phi_0$ and particle size relative to the drop radius $\epsilon = a/R$.
We start by varying the particle properties while keeping $\mathcal{D} = 0.10$ and $\kappa = 0.01$, which roughly represent the values in our experiments.
We solve the concentration field in each phase numerically, and assess the effects of the coating on the depletion dynamics by examining the temporal evolution of the total mass in the bulk $N_{\rm b}(t)$.

\subsection{Numerical predictions and transport regimes}

\begin{figure*}[t]
    \centering
    \includegraphics[width=\linewidth]{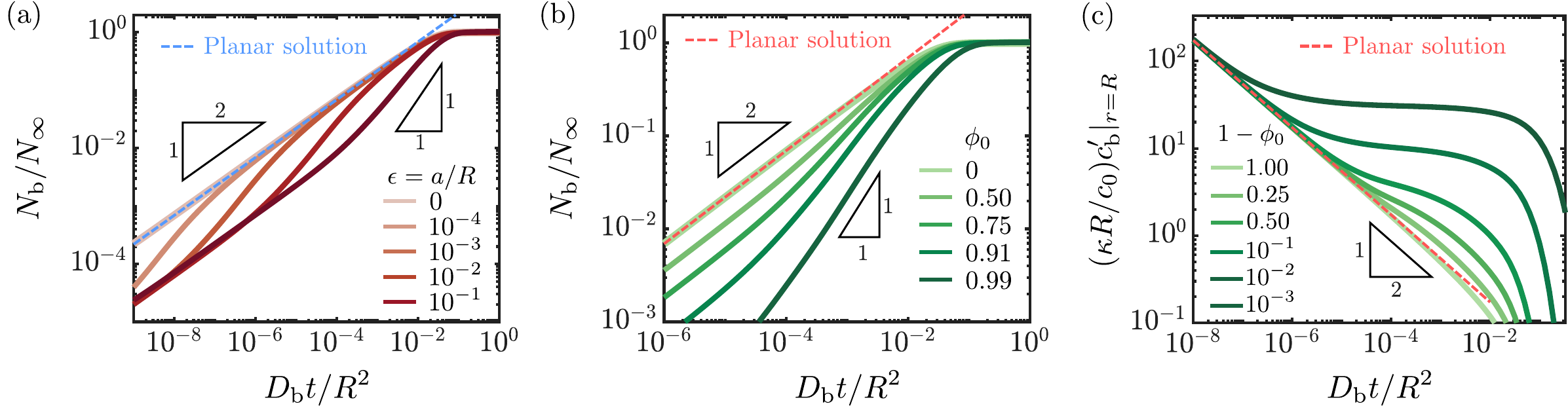}
    \caption{Numerical results for the diffusion from a coated drop. (a) The normalized total mass per unit length $N_{\rm b}/N_\infty$ against the dimensionless time $D_{\rm b}t/R^2$ for a coated drop with $\phi_0 = 0.91$ and varying the relative particle size $\epsilon = a/R$.
    At early times, the bare drop follows the planar solution \eqref{eq:n_1D} (dotted line), while the coated drops are shifted downwards still following the $N_{\rm b} \sim t^{1/2}$ scaling. 
    (b) Keeping the particle size constant $\epsilon = 0.05$ and varying the coverage fraction, yields a similar behavior. 
    Yet, pushing the coverage to extreme values $\phi_0 = 0.99$ delays the depletion of the drop.
    (c) The normalized interfacial concentration gradient $(\kappa R /c_0) \left.c_{\rm b}'\right|_{r=R}$ against the dimensionless time.
    At early times the concentration gradient decreases as $1/t^{1/2}$.
    When the surface coverage is pushed to extreme values, the gradient becomes time-independent.}
    \label{Fig5}
\end{figure*}

We compute the total mass of diffused material $N_{\rm b}(t)$ by integrating the concentration field in the bulk over the surface area available for diffusion $A(r)$.
Figure \ref{Fig5}(a) shows the total mass $N_{\rm b}(t)$ diffused into the bulk against dimensionless time $D_{\rm b}t/R^2$.
The total mass has been scaled by the equilibrium value $N_\infty$, which becomes slightly lower than $\pi R^2 c_0$ due to the volume occupied by the particles.
For an uncoated interface ($a/R = 0$) the total mass initially scales as $N_{\rm b} \sim t^{1/2}$.
The exact solution at early times can be solved analytically by neglecting the curvature terms in \eqref{eq:diffusion_o}, leading to~$c_{\rm b}(r,t) = c_0 (\sqrt{\mathcal{D}}+\kappa)^{-1} {\rm Erfc}[(r-R)/\sqrt{4D_{\rm b}t}]$ \cite{crank1979mathematics}, where ${\rm Erfc}$ is the complementary error function. 
Integrating this concentration field in the bulk gives the total mass for a planar uncoated interface
\begin{equation}
N_{\rm pl}(t) =  \frac{\pi R^2 c_0}{\sqrt{\mathcal{D}}+ \kappa} \left(\frac{4D_{\rm b}t}{\pi R^2} \right)^{1/2},
\label{eq:n_1D}
\end{equation}
which perfectly matches the early time behavior of a bare drop (dashed line in Fig.~\ref{Fig5}(a)).
During the later stages of diffusion, the curvature and the finite amount of solute initially present become important, leading to a deviation of $N_{\rm b}(t)$ from the Cartesian solution as it approaches its equilibrium value $N_{\rm b} = \pi R^2 c_0$.

Introducing a particle layer with varying values of $\epsilon$ and $\phi_0 = 0.91$, we find that the curves for $N_{\rm b}(t)$ deviate from the planar solution.
At early times, the total mass still scales as  $N_{\rm b} \sim t^{1/2}$, but with a pre-factor lower than the bare drop.
Each curve undergoes a transition, where the total mass now grows linearly with time $N_{\rm b} \sim t$.
This transition partially compensates for the initial reduction in flux, so that each curve eventually approaches the bare-drop solution.
The times at which these transitions occur increase with $\epsilon$.
For sufficiently small particle sizes, the curves exhibit the behavior predicted by \eqref{eq:n_1D} at late times, while for large particle sizes equilibrium is reached before the transition can take place.
Despite these differences, all curves reach the equilibrium value at approximately the same time.

We next examining the total mass, keeping the particle size $\epsilon = 0.05$ fixed and varying the extent of the surface coverage $\phi_0$ (Fig.~\ref{Fig5}(b)).
The effect of the particle layer is similar, as each curve again falls below the solution of the bare drop.
At early times, the curves follow the same $N_{\rm b} \sim t^{1/2}$ scaling behavior, but with a pre-factor that decreases with $\phi_0$.
The curves again transition to the intermediate regime; yet the linear growth $N_{\rm b} \sim t$ becomes apparent only for large surface coverages.
At later times, the curves fall on top of the bare solution, reaching the equilibrium value $\pi R^2 c_0$ at the same time.
Only when the surface coverage is increased to the extreme value $\phi_0 = 0.99$ does the depletion time increase by roughly a factor of two. 
Therefore, the particles appear to have a subtle role in the diffusion dynamics.
The coating strongly affects the transport at early times, with the total mass undergoing a series of different temporal dynamics.
Yet, the time it takes to fully deplete the drop remains unaffected, unless the surface coverage is increased to values beyond the close-packing limit.
The remaining question is to quantify how the surface coverage dictates the different regimes of the total mass of diffused material.

The three temporal regimes arise from how the particle layer modifies the interfacial mass flux $J =  A(r) D_{\rm b}  c_{\rm b}'|_{r=R}$.
Because the total mass satisfies $N_{\rm b} = \int J\,{\rm d t}$, the evolution of $N_{\rm b}$ is governed by the interfacial concentration gradient.
At the onset of diffusion, the characteristic concentration scale in the bulk becomes $c_{\rm b} \sim c_0/(\sqrt{\mathcal{D}} + \kappa)$ and develops over a length scale $(D_{\rm b}t)^{1/2}$.
The resulting concentration gradient thus scales as $c_{\rm b}' \sim [c_0/(\sqrt{\mathcal{D}} + \kappa)]/(D_{\rm b} t)^{1/2}$, which decreases with time.
As time progresses, the concentration diffuses beyond the narrow throat formed by adjacent particles at $r \approx R$, marking the onset of the intermediate regime.
In the limit of $\phi_0 \to 1$, integrating over the available area reveals that the concentration field now develops over a length scale $a\sqrt{1 - \phi_0}$.
The concentration gradient then follows the scaling $c_{\rm b}' \sim [c_0/(\mathcal{D} + \kappa)]/(a\sqrt{1 - \phi_0})$, which is time-independent. 
Finally, at the later stages, the concentration within the layer becomes constant and the effective flux is controlled by regions where the local coverage vanishes $\phi(r) \approx 0$.
Thus, diffusion continues as if the particle layer were absent, leading to a concentration gradient that regains the form $c_{\rm b}' \sim [c_0/(\sqrt{\mathcal{D}} + \kappa)]/(D_{\rm b} t)^{1/2}$ over an uncoated area $A(r) \sim A_0$.

To test these scaling behaviors, we plot in Fig.~\ref{Fig5}(c) the scaled concentration gradient $(\kappa R/c_0) c_{\rm b}'(r=R,t)$ against the dimensionless time $D_{\rm b}t/R^2$ obtained from the finite difference simulations.
The curves confirm the scaling behavior expected for the different temporal regimes.
Provided the surface coverage is not too high, the concentration gradient follows the planar $t^{-1/2}$ behavior at early and late times, separated by a brief intermediate regime.
For sufficiently high surface coverages, the late time regime is never reached, as the concentration gradient abruptly vanishes, signaling the drop depletion.
However, the intermediate regime becomes much more pronounced and prolonged in the limit $1 -\phi_0 \to 0$, during which the concentration gradient indeed becomes time-independent and scales as $c_{\rm b}' \sim [c_0/(\mathcal{D} + \kappa)]/(a\sqrt{1 - \phi_0})$.

Having established the temporal evolution of the concentration gradient, we now return to the total mass.
Injecting the different behaviors of the mass flux in the integral $N_{\rm b} = \int J\,{\rm d t}$, the total mass adopts the following scaling behaviors
\begin{equation}
N_{\rm b} \sim 
\begin{cases}
	\displaystyle\frac{R^2 c_0}{\sqrt{\mathcal{D}} + \kappa} (1-\phi_0)\left(\frac{D_{\rm b}t}{R^2}\right)^{1/2}, & {\rm for} \quad t \ll \displaystyle \frac{\beta a^2 (1-\phi_0)}{D_{\rm b}}, \\[12pt]
	\displaystyle\frac{R^2 c_0}{\mathcal{D} + \kappa} \frac{(1-\phi_0)^{1/2}R}{a}\left(\frac{D_{\rm b}t}{R^2}\right), & {\rm for} \quad \displaystyle \frac{\beta a^2 (1-\phi_0)}{D_{\rm b}} \ll t \ll \displaystyle \frac{\beta a^2}{D_{\rm b} (1-\phi_0)}, \\[12pt]
	\displaystyle\frac{R^2 c_0}{\sqrt{\mathcal{D}} + \kappa}\left(\frac{D_{\rm b}t}{R^2}\right)^{1/2}, & {\rm for} \quad t \gg \displaystyle \frac{\beta a^2}{D_{\rm b} (1-\phi_0)}, \\[12pt]
\end{cases}
\label{eq:tau_dep}
\end{equation}
which confirm the trends observed in Figs.~\ref{Fig5}(a,b).
Here, we have defined the parameter $\beta = [(\mathcal{D} + \kappa)/(\sqrt{\mathcal{D}} + \kappa)]^2$, which depends exclusively on the transport and thermodynamic properties.
With the exception of $\beta$, the time scales that define each regime adopt the same functional form as the planar interface \cite{van2025scaling}; yet, sufficiently coated cylindrical interfaces may not exhibit the late regime of a bare drop characterized by \eqref{eq:n_1D}.
Instead, the intermediate regime becomes so prolonged that the drop gets depleted before the secondary transition can take place.
Therefore, a comparison between the depletion time of a bare drop and the duration of the intermediate regime can provide an indication as to when the layer will hinder mass transport.

\subsection{Criteria for transport hindrance}

\begin{figure*}[t]
    \centering
    \includegraphics[width=\linewidth]{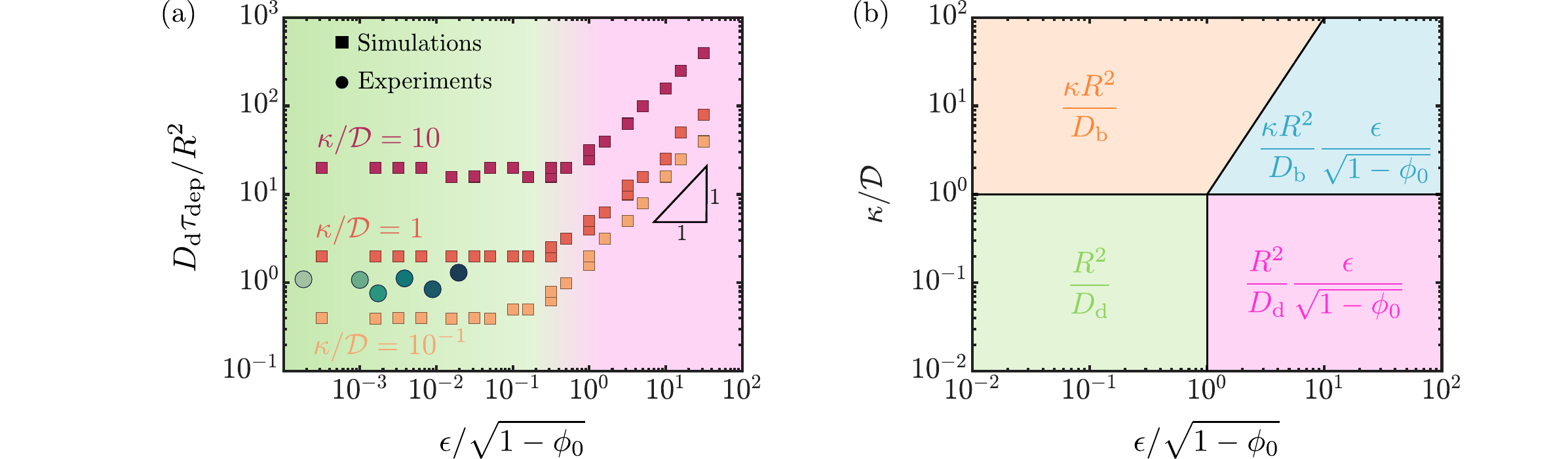}
    \caption{Phase plots for the depletion dynamics of coated drops.
    (a) Plotting scaled depletion time $\tau_{\rm dep} R^2/D_{\rm d}$ against the coating parameter $a/(R\sqrt{1-\phi_0})$, two limiting behaviors emerge for constant values of $\kappa/\mathcal{D}$. For low values of $a/(R\sqrt{1-\phi_0})$, the depletion time remains constant; yet, for large values of $a/(R\sqrt{1-\phi_0})$, the depletion time scales linearly with $a/(R\sqrt{1-\phi_0})$.
    (b) Phase diagram showing the different asymptotic cases for the depletion time as a function of the coating and transport properties. 
    On the horizontal axis the coating properties quantified by the parameter $a/(R\sqrt{1-\phi_0})$, while on the vertical axis the transport properties are quantified by the parameter $\kappa/\mathcal{D}$.}
    \label{Fig6}
\end{figure*}

We recall that for a bare drop, the depletion time adopts different scaling behaviors, depending on the transport properties.
For $\kappa \ll \sqrt{\mathcal{D}}$, transport is limited by the diffusivity of the drop and $\tau_{\rm dep} \sim R^2/D_{\rm d}$, which can be found by setting $N_{\rm b} = \pi R^2 c_0$ in $\eqref{eq:n_1D}$.
Conversely, when $\kappa \gg \mathcal{D}$ and $\kappa \gg 1$, transport is limited by the poor solubility in the bulk and the concentration difference within the drop becomes of order $\sqrt{\mathcal{D}}/\kappa$.
Because this value is very small, the drop concentration becomes uniform very fast and \eqref{eq:n_1D} is no longer valid.
In this limit, the problem can be reduced to bulk diffusion with a quasi-steady interfacial concentration.
Diffusion in the bulk occurs much faster than the rate at which the drop concentration decreases, reminiscent of bubble and drop dissolution \cite{epstein1950stability,duncan2004test}.
Even though this simplified problem cannot be solved analytically for a cylindrical drop  \cite{penas2017diffusion}, the depletion time can be estimated to scale as $\tau_{\rm dep} \sim \kappa R^2/D_{\rm b} \sim (\kappa/\mathcal{D}) R^2/D_{\rm d}$.
Finally, we note that a third case exists for $\sqrt{\mathcal{D}}\ll \kappa \ll 1$, where depletion is limited by the diffusivity of the bulk.
Setting $N_{\rm b}= \pi R^2 c_0$ in $\eqref{eq:n_1D}$ for $\sqrt{\mathcal{D}}\ll \kappa$, leads to a depletion time that scales as $\tau_{\rm dep} \sim \kappa^2 R^2/D_{\rm b} \sim (\kappa^2/\mathcal{D}) R^2/D_{\rm d}$.
The key question is whether the particle-dominated intermediate regime persists long enough to affect the overall depletion time.

For drop-limited depletion $\kappa \ll \sqrt{\mathcal{D}}$, the comparison of time scales reveals that particles will delay mass transfer when $\epsilon/\sqrt{1-\phi_0} \gtrsim 1$.
The resulting particle-dominated mass flux leads to a depletion time that scales as $\tau_{\rm dep} \sim (R^2/D_{\rm d})\epsilon/\sqrt{1-\phi_0} $.
For solubility-limited depletion $\kappa \gg \mathcal{D}$ and $\kappa\gg 1$, the criterion for transport hindrance becomes $\epsilon/\sqrt{1-\phi_0} \gtrsim \sqrt{\kappa}$, with the depletion time scaling as $\tau_{\rm dep} \sim (\kappa R^2/D_{\rm b})\epsilon/\sqrt{1-\phi_0} $.
Finally, for bulk-limited depletion $\sqrt{\mathcal{D}}\ll \kappa \ll 1$, the depletion time adopts the same scaling, but occurs instead when $\epsilon/\sqrt{1-\phi_0} \gtrsim \kappa$.

In all three cases, the coating amplifies the depletion time from the bare drop by a factor $\epsilon/\sqrt{1-\phi_0} $.
Thus, the two particle properties combine to form a single dimensionless parameter that quantifies the coating effects. 
To test this modified scaling behavior, we compute the depletion time over a wide range of values of $\epsilon$, $\phi_0$, $\kappa$, and $\mathcal{D}$.
Figure \ref{Fig6}(a) shows the depletion time $\tau_{\rm dep}$, now scaled by the drop's diffusive time scale $R^2/D_{\rm d}$, against the coating parameter $\epsilon/\sqrt{1-\phi_0} $.
For constant values of $\kappa/\mathcal{D}$, the depletion time remains constant for $\epsilon/\sqrt{1-\phi_0}\ll 1$, while in the opposite limit it scales linearly with $\epsilon/\sqrt{1-\phi_0}$.
These results confirm our scaling predictions for how the coating modifies the depletion time for limiting values of the coating parameter.
To place our experimental results within this context, we overlay the average of our experimentally-computed depletion times of Fig.~\ref{Fig3}(c).
The experimental data are consistent with the scaling predictions and lie in the regime where the particles have essentially no effect on how fast depletion occurs.

To combine all the different cases that characterize the depletion dynamics of a coated drop, we present the asymptotic limits in a phase diagram in Fig.~\ref{Fig6}(b).
The horizontal axis indicates the coating properties, which can be quantified by the dimensionless parameter $\epsilon/\sqrt{1-\phi_0}$.
Similarly, the transport and thermodynamic properties can also be grouped into the parameter $\kappa/\mathcal{D}$, which is used as the vertical axis.
For $\kappa/\mathcal{D} < 1$, transport is limited by the drop's diffusivity, becoming further restricted by the particle coating when $\epsilon/\sqrt{1-\phi_0} > 1$.
For $\kappa/\mathcal{D} > 1$, the depletion is now governed by the bulk solubility, with the particles further prolonging depletion when $\epsilon/\sqrt{1-\phi_0} > \sqrt{\kappa} $.
We note that the present phase does not account for the regime limited by the bulk's diffusivity, as $\kappa$ and $\mathcal{D}$ govern the transport dynamics separately.

\section{Conclusion \& Outlook}
We have investigated how particles adhered to interfaces affect interfacial diffusion from two-dimensional drops.
Using fluorescent microscopy experiments, we demonstrated that the particles have a minimal effect on the transport dynamics, as quantified by the penetration length and depletion time.
These results were rationalized by a one-dimensional diffusion model consisting of a particle-coated interface. 
The model revealed that the presence of particles induced three temporal transport regimes, each governed by the local concentration gradient within the layer.
A comparison between the particle-dominated regime and the depletion time of the drop indicated when the coating delays mass transport, consistent with our experimental observations.
The resulting criterion involves a single dimensionless parameter $\epsilon/\sqrt{1-\phi_0}$, which combines the relative particle size $\epsilon =a/R$ and surface coverage $\phi_0$.
Regardless of the transport properties, $\epsilon/\sqrt{1-\phi_0}$ must be large for the particle coating to have any effect on the transport dynamics.
Even though the particle radius can be controlled, $\epsilon$ is typically limited to values below $\sim 0.1$.
Therefore, transport can be significantly limited in practice via the surface coverage $\phi_0$.
For a monolayer of spherical particles, the highest attainable fraction corresponds to hexagonal close-packing $\phi_0 = 0.91$, which is insufficient to strongly hinder transport, even for large particle sizes.
To sufficiently prolong depletion, $\phi_0$ needs to exceed the hexagonal close-packing limit, which can be achieved via thermal \cite{yow2009release} or chemical \cite{thompson2010covalently} treatment of the particle layer.

More broadly, these results clarify how particle-coated interfaces regulate mass transfer in Pickering emulsions used in food products \cite{gunes2017oleofoams,sarkar2020sustainable}, controlled release \cite{dinsmore2002colloidosomes}, and multiphase chemical conversion \cite{ruiz2022catalysis}. 
In these systems, particle layers are often assumed to significantly hinder transport; yet, our results show that their effects on transport remains weak in practical settings. 
This behavior is specific to discrete microscopic particle coatings, where transport occurs through pores formed between adjacent particles, and differs from surfactant or lipid layers, where interfacial transport is governed by mechanisms on the molecular scale. 
Bridging these length scales may provide a unified framework to understand how interfacial coatings govern mass transfer.

\section*{Acknowledgements}
We acknowledge B.P. Binks, A.G. Marin, and E. Sharma for helpful discussions. 
We also thank S. Shruti for experimental assistance.
This work is supported by the Dutch Research Council (NWO) under grant number 19929.

%

\bibliographystyle{apsrev4-2}

\bibliography{refs}

@article{sagis2011dynamic,
  title={Dynamic properties of interfaces in soft matter: Experiments and theory},
  author={Sagis, Leonard MC},
  journal={Rev. Mod. Phys.},
  volume={83},
  number={4},
  pages={1367--1403},
  year={2011},
  publisher={APS}
}

@article{cates2018theories,
  title={Theories of binary fluid mixtures: from phase-separation kinetics to active emulsions},
  author={Cates, Michael E and Tjhung, Elsen},
  journal={J. Fluid Mech.},
  volume={836},
  pages={P1},
  year={2018},
  publisher={Cambridge University Press}
}

@article{chevalier2013emulsions,
  title={Emulsions stabilized with solid nanoparticles: Pickering emulsions},
  author={Chevalier, Yves and Bolzinger, Marie-Alexandrine},
  journal={Colloids Surf., A: Physicochem. Eng.
Aspects},
  volume={439},
  pages={23--34},
  year={2013},
  publisher={Elsevier}
}

@article{yan2020protein,
  title={Protein-stabilized Pickering emulsions: Formation, stability, properties, and applications in foods},
  author={Yan, Xiaojia and Ma, Cuicui and Cui, Fengzhan and McClements, David Julian and Liu, Xuebo and Liu, Fuguo},
  journal={Trends Food Sci. Technol.},
  volume={103},
  pages={293--303},
  year={2020},
  publisher={Elsevier}
}

@article{tang2015stimuli,
  title={Stimuli-responsive Pickering emulsions: recent advances and potential applications},
  author={Tang, Juntao and Quinlan, Patrick James and Tam, Kam Chiu},
  journal={Soft Matter},
  volume={11},
  number={18},
  pages={3512--3529},
  year={2015},
  publisher={Royal Society of Chemistry}
}

@article{crossley2010solid,
  title={Solid nanoparticles that catalyze biofuel upgrade reactions at the water/oil interface},
  author={Crossley, Steven and Faria, Jimmy and Shen, Min and Resasco, Daniel E},
  journal={Science},
  volume={327},
  number={5961},
  pages={68--72},
  year={2010},
  publisher={American Association for the Advancement of Science}
}

@article{rodriguez2020catalysis,
  title={Catalysis in Pickering emulsions},
  author={Rodriguez, Ana Maria Bago and Binks, Bernard P},
  journal={Soft Matter},
  volume={16},
  number={45},
  pages={10221--10243},
  year={2020},
  publisher={Royal Society of Chemistry}
}

@article{dinsmore2002colloidosomes,
  title={Colloidosomes: selectively permeable capsules composed of colloidal particles},
  author={Dinsmore, AD and Hsu, Ming F and Nikolaides, MG and Marquez, Manuel and Bausch, AR and Weitz, DA},
  journal={Science},
  volume={298},
  number={5595},
  pages={1006--1009},
  year={2002},
  publisher={American Association for the Advancement of Science}
}

@article{thompson2015colloidosomes,
  title={Colloidosomes: Synthesis, properties and applications},
  author={Thompson, Kate L and Williams, Mark and Armes, Steven P},
  journal={J. Colloid Int. Sci.},
  volume={447},
  pages={217--228},
  year={2015},
  publisher={Elsevier}
}

@article{sjoo2015barrier,
  title={Barrier properties of heat treated starch Pickering emulsions},
  author={Sj{\"o}{\"o}, Malin and Emek, Sinan Cem and Hall, Tina and Rayner, Marilyn and Wahlgren, Marie},
  journal={J. Colloid Int. Sci.},
  volume={450},
  pages={182--188},
  year={2015},
  publisher={Elsevier}
}

@article{abkarian2007dissolution,
  title={Dissolution arrest and stability of particle-covered bubbles},
  author={Abkarian, Manouk and Subramaniam, Anand Bala and Kim, Shin-Hyun and Larsen, Ryan J and Yang, <? format?> Seung-Man and Stone, Howard A},
  journal={Phys. Rev. Lett.},
  volume={99},
  number={18},
  pages={188301},
  year={2007},
  publisher={APS}
}

@article{gallo2021particle,
  title={How particle--particle and liquid--particle interactions govern the fate of evaporating liquid marbles},
  author={Gallo, A and Tavares, F and Das, Ratul and Mishra, Himanshu},
  journal={Soft Matter},
  volume={17},
  number={33},
  pages={7628--7644},
  year={2021},
  publisher={Royal Society of Chemistry}
}

@article{saczek2024impact,
  title={Impact of coating particles on liquid marble lifetime: reactor engineering approach to evaporation},
  author={Saczek, Joshua and Murphy, Koren and Zivkovic, Vladimir and Putranto, Aditya and Pramana, Stevin S},
  journal={Soft Matter},
  volume={20},
  number={29},
  pages={5822--5835},
  year={2024},
  publisher={Royal Society of Chemistry}
}

@article{laborie2013coatings,
  title={How coatings with hydrophobic particles may change the drying of water droplets: incompressible surface versus porous media effects},
  author={Laborie, Benoit and Lachauss{\'e}e, Florent and Lorenceau, Elise and Rouyer, Florence},
  journal={Soft Matter},
  volume={9},
  number={19},
  pages={4822--4830},
  year={2013},
  publisher={Royal Society of Chemistry}
}

@article{kim2007colloidal,
  title={Colloidal assembly route for responsive colloidosomes with tunable permeability},
  author={Kim, Jin-Woong and Fern{\'a}ndez-Nieves, Alberto and Dan, Nily and Utada, Andrew S and Marquez, Manuel and Weitz, David A},
  journal={Nano Letters},
  volume={7},
  number={9},
  pages={2876--2880},
  year={2007},
  publisher={ACS Publications}
}

@article{yow2009release,
  title={Release profiles of encapsulated actives from colloidosomes sintered for various durations},
  author={Yow, Huai Nyin and Routh, Alexander F},
  journal={Langmuir},
  volume={25},
  number={1},
  pages={159--166},
  year={2009},
  publisher={ACS Publications}
}

@article{li2025pickering,
  title={Pickering emulsions with low interface coverage but enhanced stability for emulsion interface catalysis and SERS-based detection},
  author={Li, Mingkun and Song, Qing and Wang, Yilin and Liu, Bing},
  journal={Nat. Commun.},
  volume={16},
  number={1},
  pages={2490},
  year={2025},
  publisher={Nature Publishing Group UK London}
}

@article{thompson2010covalently,
  title={Covalently cross-linked colloidosomes},
  author={Thompson, Kate Louise and Armes, SP and Howse, JR and Ebbens, S and Ahmad, I and Zaidi, JH and York, DW and Burdis, JA},
  journal={Macromolecules},
  volume={43},
  number={24},
  pages={10466--10474},
  year={2010},
  publisher={ACS Publications}
}

@article{liu2024diffusion,
  title={Diffusion across particle-laden interfaces in Pickering droplets},
  author={Liu, Yanyan and Xu, Mingjun and Portela, Luis M and Garbin, Valeria},
  journal={Soft Matter},
  volume={20},
  number={1},
  pages={94--102},
  year={2024},
  publisher={Royal Society of Chemistry}
}

@article{penas2017diffusion,
  title={Diffusion of dissolved CO 2 in water propagating from a cylindrical bubble in a horizontal Hele-Shaw cell},
  author={Pe{\~n}as-L{\'o}pez, Pablo and Van Elburg, Benjamin and Parrales, Miguel A and Rodr{\'\i}guez-Rodr{\'\i}guez, Javier},
  journal={Phys. Rev. Fluids},
  volume={2},
  number={6},
  pages={063602},
  year={2017},
  publisher={APS}
}

@article{lee2008double,
  title={Double emulsion-templated nanoparticle colloidosomes with selective permeability},
  author={Lee, Daeyeon and Weitz, David A},
  journal={Advanced Materials},
  volume={20},
  number={18},
  pages={3498--3503},
  year={2008},
  publisher={WILEY-VCH Verlag Weinheim}
}

@book{crank1979mathematics,
  title={The mathematics of diffusion},
  author={Crank, John},
  year={1979},
  publisher={Oxford university press}
}

@book{bruus2007theoretical,
  title={Theoretical microfluidics},
  author={Bruus, Henrik},
  volume={18},
  year={2007},
  publisher={Oxford university press}
}

@article{van2025scaling,
  title={Scaling regimes for unsteady diffusion across particle-stabilized fluid interfaces},
  author={van Overveld, TJJM and Garbin, V},
  journal={Phys. Rev. Fluids},
  volume={10},
  number={11},
  pages={L112501},
  year={2025},
  publisher={APS}
}

@article{duncan2004test,
  title={Test of the Epstein- Plesset Model for gas microparticle dissolution in aqueous media: effect of surface tension and gas undersaturation in solution},
  author={Duncan, P Brent and Needham, David},
  journal={Langmuir},
  volume={20},
  number={7},
  pages={2567--2578},
  year={2004},
  publisher={ACS Publications}
}

@article{epstein1950stability,
  title={On the stability of gas bubbles in liquid-gas solutions},
  author={Epstein, Paul S and Plesset, Milton S},
  journal={J. Chem. Phys.},
  volume={18},
  number={11},
  pages={1505--1509},
  year={1950}
}

@incollection{jacobs1967diffusion,
  title={Diffusion processes},
  author={Jacobs, Merkel Henry},
  booktitle={Ergebnisse der Biologie},
  pages={1--160},
  year={1967},
  publisher={Springer}
}

@article{zwanzig1992diffusion,
  title={Diffusion past an entropy barrier},
  author={Zwanzig, Robert},
  journal={J. Phys. Chem.},
  volume={96},
  number={10},
  pages={3926--3930},
  year={1992},
  publisher={ACS Publications}
}

@article{binks2002particles,
  title={Particles as surfactants—similarities and differences},
  author={Binks, Bernard P},
  journal={Curr. Opin. Colloid Interface Sci.},
  volume={7},
  number={1-2},
  pages={21--41},
  year={2002},
  publisher={Elsevier}
}

@article{pawar2011arrested,
  title={Arrested coalescence in Pickering emulsions},
  author={Pawar, Amar B and Caggioni, Marco and Ergun, Roja and Hartel, Richard W and Spicer, Patrick T},
  journal={Soft Matter},
  volume={7},
  number={17},
  pages={7710--7716},
  year={2011},
  publisher={Royal Society of Chemistry}
}

@article{dedovets2022multiphase,
  title={Multiphase microreactors based on liquid--liquid and gas--liquid dispersions stabilized by colloidal catalytic particles},
  author={Dedovets, Dmytro and Li, Qingyuan and Leclercq, Lo{\"\i}c and Nardello-Rataj, Veronique and Leng, Jacques and Zhao, Shuangliang and Pera-Titus, Marc},
  journal={Angew. Chem. Int. Ed.},
  volume={61},
  number={4},
  pages={e202107537},
  year={2022},
  publisher={Wiley Online Library}
}

@article{chang2021recent,
  title={Recent developments in catalysis with Pickering Emulsions},
  author={Chang, Fuqiang and Vis, Carolien M and Ciptonugroho, Wirawan and Bruijnincx, Pieter CA},
  journal={Green Chem.},
  volume={23},
  number={7},
  pages={2575--2594},
  year={2021},
  publisher={Royal Society of Chemistry}
}

@article{berton2015pickering,
  title={Pickering emulsions for food applications: background, trends, and challenges},
  author={Berton-Carabin, Claire C and Schro{\"e}n, Karin},
  journal={Annu. Rev. Food Sci. Technol.},
  volume={6},
  number={1},
  pages={263--297},
  year={2015},
  publisher={Annual Reviews}
}

@article{bais2005rheological,
  title={Rheological characterization of polysaccharide--surfactant matrices for cosmetic O/W emulsions},
  author={Bais, D and Trevisan, A and Lapasin, ROMANO and Partal, P and Gallegos, C},
  journal={J. Colloid Int. Sci.},
  volume={290},
  number={2},
  pages={546--556},
  year={2005},
  publisher={Elsevier}
}

@article{sun2019coated,
  title={Coated colloidosomes as novel drug delivery carriers},
  author={Sun, Qian and Chen, Jian-Feng and Routh, Alexander F},
  journal={Expert Opin. Drug Deliv.},
  volume={16},
  number={9},
  pages={903--906},
  year={2019},
  publisher={Taylor \& Francis}
}

@article{thijssen2018interfacial,
  title={Interfacial rheology of model particles at liquid interfaces and its relation to (bicontinuous) Pickering emulsions},
  author={Thijssen, Job HJ and Vermant, Jan},
  journal={Journal of Physics: Condensed Matter},
  volume={30},
  number={2},
  pages={023002},
  year={2018},
  publisher={IOP Publishing}
}

@article{van2017interfacial,
  title={Interfacial rheology of sterically stabilized colloids at liquid interfaces and its effect on the stability of pickering emulsions},
  author={Van Hooghten, Rob and Blair, Victoria E and Vananroye, Anja and Schofield, Andrew B and Vermant, Jan and Thijssen, Job HJ},
  journal={Langmuir},
  volume={33},
  number={17},
  pages={4107--4118},
  year={2017},
  publisher={ACS Publications}
}

@article{hunter2008role,
  title={The role of particles in stabilising foams and emulsions},
  author={Hunter, Timothy N and Pugh, Robert J and Franks, George V and Jameson, Graeme J},
  journal={Advances in colloid and interface science},
  volume={137},
  number={2},
  pages={57--81},
  year={2008},
  publisher={Elsevier}
}

@article{loussert2016drying,
  title={Drying dynamics of a charged colloidal dispersion in a confined drop},
  author={Loussert, Charles and Bouchaudy, Anne and Salmon, Jean-Baptiste},
  journal={Physical Review Fluids},
  volume={1},
  number={8},
  pages={084201},
  year={2016},
  publisher={APS}
}

@article{milark2026confined,
  title={Confined drying of a binary liquid mixture droplet: A quantitative interferometric study under humidity control},
  author={Milark, Ole and Salmon, Jean-Baptiste and Sobac, Benjamin},
  journal={Physical Review Fluids},
  volume={11},
  number={3},
  pages={033603},
  year={2026},
  publisher={APS}
}

@article{clement2004evaporation,
  title={Evaporation of liquids and solutions in confined geometry},
  author={Cl{\'e}ment, F and Leng, J},
  journal={Langmuir},
  volume={20},
  number={16},
  pages={6538--6541},
  year={2004},
  publisher={ACS Publications}
}

@article{boulogne2013buckling,
  title={The buckling and invagination process during consolidation of colloidal droplets},
  author={Boulogne, Fran{\c{c}}ois and Giorgiutti-Dauphin{\'e}, Fr{\'e}d{\'e}rique and Pauchard, Ludovic},
  journal={Soft Matter},
  volume={9},
  number={3},
  pages={750--757},
  year={2013},
  publisher={Royal Society of Chemistry}
}

@article{reguera2001kinetic,
  title={Kinetic equations for diffusion in the presence of entropic barriers},
  author={Reguera, David and Rubi, JM},
  journal={Phys. Rev. E},
  volume={64},
  number={6},
  pages={061106},
  year={2001},
  publisher={APS}
}

@article{kalinay2006corrections,
  title={Corrections to the Fick-Jacobs equation},
  author={Kalinay, P and Percus, JK},
  journal={Physical Review E—Statistical, Nonlinear, and Soft Matter Physics},
  volume={74},
  number={4},
  pages={041203},
  year={2006},
  publisher={APS}
}

@article{gunes2017oleofoams,
  title={Oleofoams: Properties of Crystal-Coated Bubbles from Whipped Oleogels Evidence for Pickering Stabilization},
  author={Gunes, DZ and Murith, Mathieu and Godefroid, Julie and Pelloux, Cindy and Deyber, H{\'e}l{\`e}ne and Schafer, O and Breton, O},
  journal={Langmuir},
  volume={33},
  number={6},
  pages={1563--1575},
  year={2017},
  publisher={ACS Publications}
}

@article{sarkar2020sustainable,
  title={Sustainable food-grade Pickering emulsions stabilized by plant-based particles},
  author={Sarkar, Anwesha and Dickinson, Eric},
  journal={Current Opinion in Colloid \& Interface Science},
  volume={49},
  pages={69--81},
  year={2020},
  publisher={Elsevier}
}

@article{ruiz2022catalysis,
  title={Catalysis at the solid--liquid--liquid interface of water--oil Pickering emulsions: A tutorial review},
  author={Ruiz, M Pilar and Faria, Jimmy A},
  journal={ACS Engineering Au},
  volume={2},
  number={4},
  pages={295--319},
  year={2022},
  publisher={ACS Publications}
}

\end{document}